\documentclass[rmp,twocolumn,floatfix,superscriptaddress]{revtex4}
\RequirePackage[sort&compress]{natbib}

\usepackage{amssymb,amsfonts,amsmath}
\usepackage{bm}
\usepackage[pdftex]{graphicx}
\usepackage{xcolor}
\usepackage{setspace}
\usepackage{subfigure}

\usepackage[font=sf,size=footnotesize,justification=RaggedRight,singlelinecheck=false]{caption}
\usepackage[sf,bf,small]{titlesec}

\definecolor{darkGray}{RGB}{153,153,153}
\definecolor{darkBlue}{RGB}{37,113,161}
\definecolor{darkGreen}{RGB}{113,161,37}
\definecolor{darkRed}{RGB}{186,21,24}

\usepackage{booktabs}
\newcommand{\mytoprule}{\specialrule{0.1em}{0em}{0em}}
\newcommand{\mybottomrule}{\specialrule{0.1em}{0em}{0em}}

\usepackage{textcomp}

\newcommand{\hs}{\@~}

\bibpunct{}{}{,}{s}{,}{,}

\begin{document}
\makeatletter
\renewcommand\@biblabel[1]{#1.}
\makeatother

\title{Identifying modular flows on multilayer networks reveals highly overlapping organization in social systems}

\author{Manlio De Domenico$^{1}$, Andrea Lancichinetti$^{2}$, Alex Arenas$^{1}$, Martin Rosvall$^{2}$\\
\normalsize{$^{1}$Departament d'Enginyeria Inform\`atica i Matem\`atiques, Universitat Rovira i Virgili, 43007 Tarragona, Spain,}
\normalsize{$^{2}$Integrated Science Lab, Department of Physics, Ume{\aa} University, SE-901 87 Ume{\aa}, Sweden}
}

\begin{abstract}
Unveiling the community structure of networks is a powerful methodology to comprehend interconnected systems across the social and natural sciences. 
To identify different types of functional modules in interaction data aggregated in a single network layer, researchers have developed many powerful methods. 
For example, flow-based methods have proven useful for identifying modular dynamics in weighted and directed networks that capture constraints on flow in the systems they represent.
However, many networked systems consist of agents or components that exhibit multiple layers of interactions.
Inevitably, representing this intricate network of networks as a single aggregated network leads to information loss and may obscure the actual organization. 
Here we propose a method based on compression of network flows that can identify modular flows in non-aggregated multilayer networks.
Our numerical experiments on synthetic networks show that the method can accurately identify modules that cannot be identified in aggregated networks or by analyzing the layers separately.
We capitalize on our findings and reveal the community structure of two multilayer collaboration networks: scientists affiliated to the Pierre Auger Observatory and scientists publishing works on networks on the arXiv. Compared to conventional aggregated methods, the multilayer method reveals smaller modules with more overlap that better capture the actual organization.
\end{abstract}

\maketitle

\section*{Introduction}

Multifaceted relationships between numerous components in social and biological systems make them inherently complex to analyze \cite{scott1988social,kitano2002systems}. 
Data about these interactions have become increasingly available and network analysis have emerged as an essential tool for studying their function \cite{newman2003structure,barrat2004architecture,boccaletti2006complex}. 
For large networks, detailed modeling of individual components and their interactions is unfeasible and researchers instead seek to simplify 
and highlight important large-scale functional structures in the networks.
Depending on the system under study and the research question at hand, researchers use methods that either operate on the plain topology of the network itself \cite{barabasi1999emergence,newman2002assortative} or, to capture flow processes through the real system, on dynamics modeled on the network \cite{brin1998anatomy,vespignani2012modelling}.
In any case, an important objective is to detect so-called communities \cite{Fortunato201075}, topological groups of nodes with higher internal than external density of links compared to null models \cite{newman2006modularity,lancichinetti2011finding,peixoto2013parsimonious} or, alternatively, modules that capture flows for a relatively long time \cite{simonsen2004diffusion,rosvall2008maps,delvenne2010stability}.

However, community-detection methods generally assume that a single type of static link, weighted and directed at best, can account for all types of interactions between nodes in the network. 
This assumption oversimplifies the multifaceted nature of relationships in real systems with important consequences.
Aggregating multiple types of relationships into a single weighted and directed network can distort both the topology of the network and the dynamics on the network \cite{rosvall2013networks}.
Take social relationships as an example, where the way the same individual interacts with her relatives, friends, and colleagues may depend on location, time, or means of interaction. Is she at home or at the office? Is it weekday or weekend? Is she communicating by phone or by Facebook? If all contact events are aggregated into a single network layer, important temporal \cite{holme2012temporal} and structural \cite{buldyrev2010catastrophic} information is inevitably lost. Recently it has been shown that multilayer networks provide an effective framework to capture different types of interactions between nodes \cite{mucha2010community,nicosia2013growing,radicchi2013abrupt,cardillo2013emergence,gomez2013diffusion,dedomenico2013mathematical,dedomenico2014navigability}, including a generalization of modularity to identify groups in multilayer networks \cite{mucha2010community}. While the generalized null models of modularity are based on Laplacian dynamics \cite{mucha2010community}, they nevertheless favor topological groups with high link density \cite{lambiotte2008laplacian}, both within and between network layers \cite{petri2014temporal}.

\section*{Modules in multilayer networks: a flow-approach}

To identify modular flows on multilayer networks, here we introduce a method based on compression of network flows. The information-theoretic method generalizes the so-called map equation \cite{rosvall2008maps} for networks with memory \cite{rosvall2013networks} to take advantage of modular flows in multilayer networks. The framework generalizes straightforwardly, because the information-theoretic machinery remains the same and only the flow model changes, with memory of present layer rather than of previous step. This approach therefore suggests a natural concept of communities in multilayer networks as groups of nodes that capture flows within and across layers for a relatively long time.

We begin by describing how we model the dynamics and then introduce the multiplex map equation. We measure the performance on benchmark networks and contrast with results obtained with the generalization of modularity. Finally we analyze the modular flow dynamics on two multilayer collaboration networks. We have integrated the method in the Infomap software package available online \cite{mapequation}.

\subsection*{Flow dynamics on multilayer networks} A multilayer network is an efficient representation of a connected system of agents that may interact in different roles, at different times, or by different means, for example. For simplicity, we refer to them as different \emph{modes}. Each \emph{physical node} represents an agent, and each \emph{network layer} represents the constraints on flow among the agents in a given mode. Figure \ref{fig:schematicmultiplex} illustrates a multilayer network with three layers and four physical nodes. We use Latin letters to enumerate the physical nodes, Greek letters to enumerate the network layers, and pairs of Latin and Greek letters to identify node-layer tuples \cite{kivela2013multilayer}, corresponding to physical nodes in specific network layers, which we in the following refer to as \emph{state nodes} (Fig.\hs\ref{fig:schematicmultiplex}B). Sometimes empirical data allow us to assign weights to both intra- and inter-layer links between state nodes.
In such \emph{interconnected networks}, we have complete information to model dynamics with a random walker that follows links proportional to their weights within and between network layers. 
In general, with intra-layer adjacency matrix $W_{ij}^{\beta}$ of layer $\beta$ and inter-layer adjacency matrix $D_{i}^{\alpha\beta}$ of physical node $i$, the transition probabilities are
\begin{align}
\mathcal{P}_{ij}^{\alpha\beta} = \frac{D_{i}^{\alpha\beta}}{S_{i}^{\alpha}}\frac{W_{ij}^{\beta}}{s_{i}^{\beta}},\label{eq:interconnectedtransitionprob}
\end{align}
where $S_{i}^{\alpha}~=~\sum_{\beta}D_{i}^{\alpha\beta}$ are the inter-layer out-strengths and $s_{i}^{\beta}~=~\sum_j W_{ij}^{\beta}$ are the intra-layer out-strengths of node $i$ in layer $\alpha$ and $\beta$, respectively \cite{dedomenico2014navigability} (see Methods). In practice, however, often data about inter-layer link weights are scarce. That is, information about the probability of switching layer is incomplete.

In absence of empirical inter-layer weights, we use random walker dynamics with \emph{relax rate} $r$ to model movements between layers. In a given step, with probability $1-r$, the random walker moves according to the intra-layer links of the state node, and with probability $r$, the constraint to move in the current layer is relaxed and the random walker moves along any link of the physical node. In this way, the random walker switches from layer $\alpha$ to layer $\beta$ with probability $s_{i}^{\beta}/S_{i}^{\alpha}$. These dynamics are described by the transition probabilities 
\begin{align}
\mathcal{P}_{ij}^{\alpha\beta}(r) =  (1-r) \delta_{\alpha\beta} \frac{W_{ij}^{\beta}}{s_{i}^{\beta}} + r \frac{W_{ij}^{\beta}}{S_{i}},\label{eq:multilayertransitionprob}
\end{align}
with $S_{i}~=~\sum_{\beta}s_{i}^{\beta}$ independent of layer.\footnote{It is worth noting that Eq.\,(\ref{eq:multilayertransitionprob}) is equivalent to Eq.\,(\ref{eq:interconnectedtransitionprob}) when $D_{i}^{\alpha\beta}~=~(1-r)\delta_{\alpha\beta}S_{i}+rs_{i}^{\beta}$ and $S_{i}^{\alpha}~=~\sum_{\beta}s_{i}^{\beta}$.}
A relaxed step on a multilayer network resembles a teleportation step in the PageRank algorithm \cite{brin1998anatomy}, which allows a random surfer to move freely to a random website and explore the full network. However, a relaxed step only frees the constraints set by the current network layer and allows the random walker to follow a link from node $i$ to node $j$ in any network layer (Fig.\hs\ref{fig:schematicmultiplex}B). Accordingly, changing the relax rate from 0 to 1 modifies the constraints on the random walker from those that force it to be stuck in disconnected network layers to those that allow it to move more freely on the fully aggregated network. In this way, we can model the important interplay between interconnected network layers.

\subsection*{Communities in multilayer networks} There are, in principle, many ways to define communities in multilayer networks \cite{mucha2010community,loe2014comparison}. The challenge is to construct an effective framework. The challenge may seem daunting, since it is still debated how to define communities in single-layer networks \cite{Fortunato201075}, and multilayer networks are inherently more complex with simultaneous and nonlinear coupling between the layers. However, by using the fact that many networks represent constraints on flow in social and biological systems, and that multilayer networks are just a more complete description of those constraints, a generalization of flow-based community detection methods follows straightforwardly.

\begin{figure}[tbhp]
\centering
\includegraphics[width=\columnwidth]{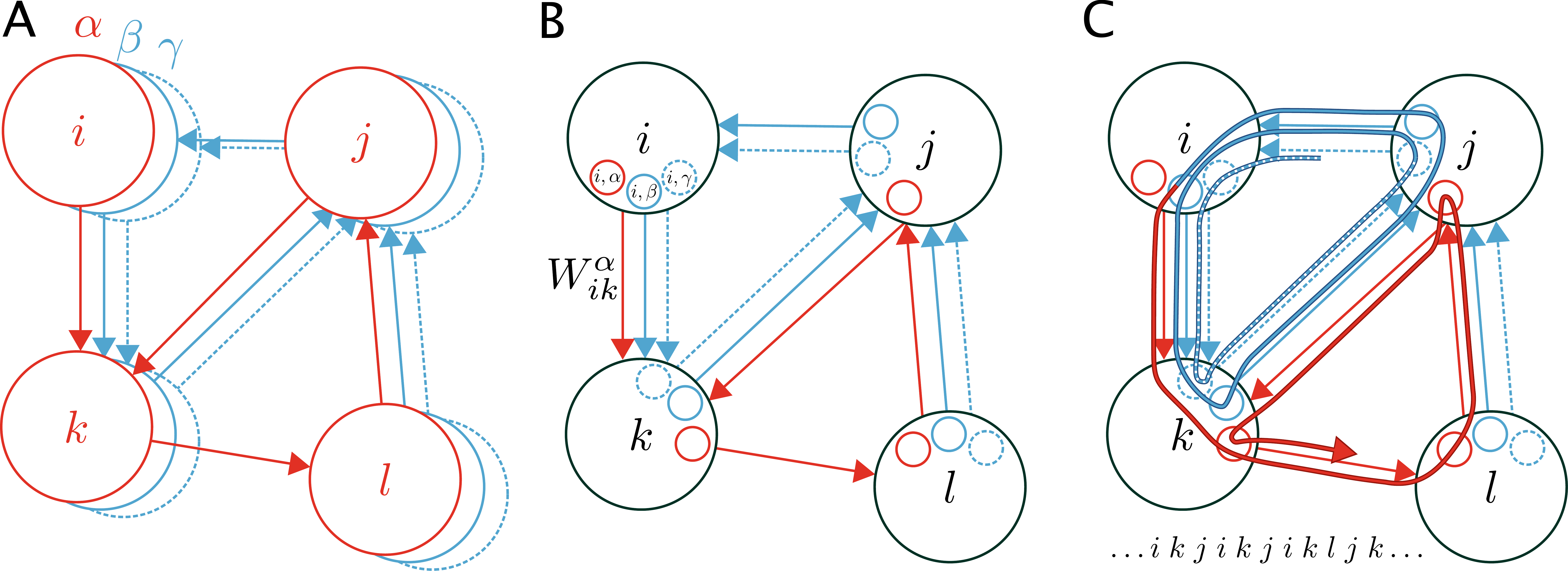}
\caption{\label{fig:schematicmultiplex}Modular flow on a multilayer network. (A) A schematic multilayer network with physical nodes $i$, $j$, $k$, and $l$ and three layers $\alpha$, $\beta$, and $\gamma$ in red, blue, and dashed blue, respectively. (B) The three layers represented as a multiplex network with physical nodes in black and state nodes $i,\alpha$ in red, blue, and dashed blue connected with intra-layer link weights $W_{ij}^{\alpha}$. (C) A random walker on the multiplex network moving between the state nodes, twice relaxing the layer constraint and following any link from the physical node of the currently visited state node (first time in $j,\gamma$ and second time in $i,\beta$).
While the random walker moves according to the weights between the state nodes, only the physical nodes are considered to be observables, as illustrated by the sequence of physical nodes that the random walker has visited. When the random walker moves along links of the red layer, it is trapped in the lower right triangle. When the random walker moves along links of the blue or dashed blue layer, it is trapped in the upper left triangle. As a consequence, the multilayer network has two overlapping modules with respect to flow.}
\end{figure}

We begin by exemplifying how we identify communities in a multilayer network. As an illustrative example, we use a social system in which nodes represent individuals and network layers represent family, friendship, and work relations, respectively. The constraints on flow in a network layer may give rise to modules with long flow persistence times. Moreover, and importantly, the modules in each network layer may or may not depend on other network layers. For example, if some friends run a business together, their module in the friendship-relations layer will correlate with their module in the work-relations layer, such that they form a single reinforced module across the two layers. Contrarily, all members of a family may not work together or even hang out as friends, such that the family module does not extend across layers. However, if some of the family members run a business together or hang out as friends, modules may overlap. That is, identifying modular flows on multilayer networks captures that individuals can belong to multiple highly interactive communities with limited information transfer between, such that information has long persistence times within communities. The schematic multilayer network in Fig.\hs\ref{fig:schematicmultiplex} illustrates. Each layer has a triangle of connected nodes that trap flow for a long time and form a module. The red network layer has very little overlap with the two identical blue network layers (Fig.\hs\ref{fig:schematicmultiplex}A). By first representing the multilayer network as a multiplex network with state nodes (Fig.\hs\ref{fig:schematicmultiplex}B), and then releasing a random walker on the multiplex network, the community structure with two overlapping modules appear (Fig.\hs\ref{fig:schematicmultiplex}C). In next the section, we make this concept of modular flow in multilayer and interconnected networks precise by generalizing the map equation.

\subsection*{The multiplex map equation}\label{sec:mapeq}

The map equation measures the length required to communicate dynamics on a network with a modular description \cite{rosvall2008maps}.
Unlike the maximum compression achieved by the entropy rate of a random walk process on the network \cite{shannon1948mathematical,schaub2012encoding}, the coding scheme of the map equation grants unique names to important structures of the network.
In this modular coding scheme, each entry to a module, and each node visit and module exit of a module, is assigned a unique code word.
With these constraints on the coding scheme, maximum compression is achieved when nodes that capture the random walk process for a relatively long time are assigned to the same module.
As a result of the duality between compressing data and finding regularities in the data, identifying the assignment of nodes into modules with maximum compression simultaneously answers: How many modules are present? And which nodes are members of which modules? 

The original map equation can be generalized by modifying the constraints on the coding scheme.
Explicitly, the three constraints on the original two-level version of the map equation for hard partitions are: (\emph{i}) A modular code structure with unique names on important structures, (\emph{ii}) movements in no more than two levels, modules and nodes, and (\emph{iii}) that each node can only belong to one module.
Constraint (\emph{i}) is essential and cannot be relaxed, but relaxing constraint (\emph{ii}) allows for multi-level solutions \cite{rosvall2011multilevel} and relaxing constraint (\emph{iii}) allows for overlapping-module solutions \cite{esquivel2011compression}.
But how should the map equation be generalized for multilayer structures?

The natural generalization is to capture the modified dynamics of the multilayer network while maintaining the essential constraint (\emph{i}) of a modular code structure with unique names on important structures.
This principle is particularly well suited to capture the fundamental notion of multilayer networks, namely that it is the very same physical object that is represented by its state nodes in each layer. Therefore, the generalization follows straightforwardly.
Whenever two state nodes that represent the same physical object are assigned to the same module, they should use a common code word (see Fig.\hs\ref{fig:schematicmultiplex}C and Methods). In colloquial terms for the example above, the colleagues who are also friends will refer to each individual by a single name that may be different from what family members use. In this way, the multilayer network modules will naturally overlap if the dynamics have such properties.
This generalization can be taken one step further by relaxing constraint (\emph{ii}) and allow for multilevel solutions with nested modules.
In fact, we have integrated both the two-level and the multilevel \emph{multiplex map equation} in the Infomap search algorithm available online \cite{mapequation}, 
but here we focus on two-level modular structures, communities.

\section*{Results and Discussion}

In this section, we first validate our framework on novel multilayer benchmark networks and then analyze two inherently multilayer collaboration networks.

\subsection*{Performance tests on multilayer benchmark networks}

To test the performance of the information-theoretic and flow-based method, we developed multilayer benchmark networks with modular structure across layers. We followed the standard approach and obtained benchmark networks from a generative model in which nodes are assigned to communities and the probability of drawing a link between two nodes depends on their community assignments \cite{girvan2002community, lancichinetti2008benchmark}. While the multiplex map equation can identify modules that independently span across any number of layers, here we consider benchmark networks with community structure in entire layers that either correlate or not. This more easily tractable scenario nevertheless highlights salient features of modular flows. As schematically illustrated in Fig.\hs\ref{fig:schematicmultiplex}, the scenario corresponds to systems that can be in different modes with dependent network layers. Using the example from above, colleagues would also be friends such that the two layers would have almost the same community structure, yet different from the community structure associated with family relations. Such redundant information is common in many social and biological networks that represent systems that can be in different modes as a whole or slowly change over time \cite{dedomenico2014compressibility}. 

Specifically, we first generated $T$ independent LFR benchmark networks \cite{lancichinetti2008benchmark} for the different modes of the system, and then sampled $L$ network layers from each of the mode networks.
We sampled the network layers by including each link with probability $1/L$ to allow a link of the mode network to be sampled once on average.
Each multilayer benchmark network thus comprises $T \times L$ layers, with $T$ sets of $L$ dependent layers.

\begin{figure}[tbhp]
\centering
\includegraphics[width=\columnwidth]{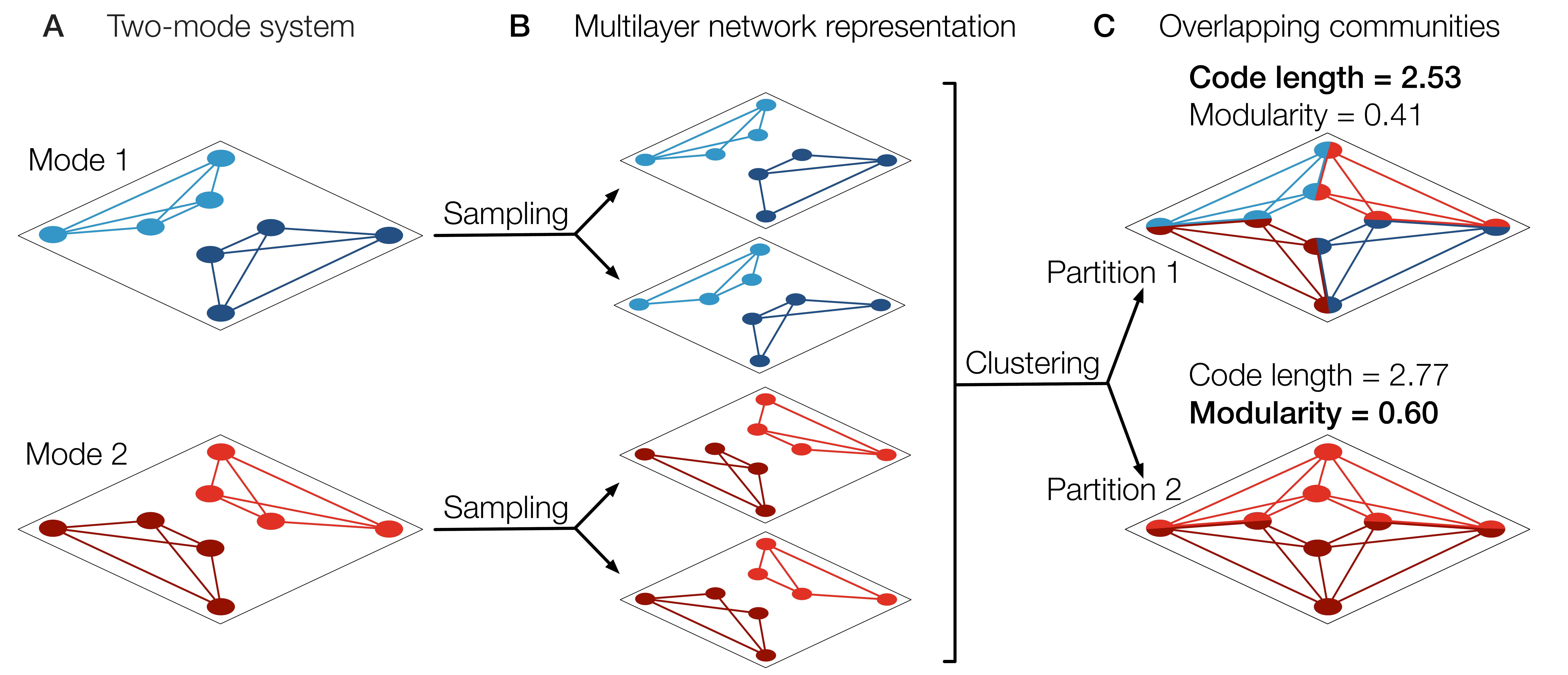}
\caption{\label{fig:schematicbenchmark}Overlapping communities in multilayer benchmark networks. We generate the multilayer networks in two steps. (A) First we generate $T$ LFR benchmark networks with well-defined communities, here illustrated with two network modes in blue and red. (B) Then we sample $L$ network layers from each mode network, here illustrated with four network layers in total. (C) Each state node in the multilayer network is classified in a community, such that communities of physical nodes map overlap. In partition 1, each state node is correctly classified. In partition 2, however, the light and dark color shades are assigned to the same module, respectively. While these communities provide the correct partition of each slice, they fail to capture the communities of the two original mode networks. Generalized modularity favors partition 2 whereas the multiplex map equation favors partition 1.}
\end{figure}

Figure \ref{fig:schematicbenchmark} schematically illustrates a multilayer benchmark network with $T=L=2$. The challenge is to reveal the community structure of each mode network, which corresponds to simultaneously reveal the community structure in each layer and identify the mode network from which the layer was sampled. To make the test realistic, we only provided the algorithm with the $T \times L$ network layers, and neither input any information about the number of mode networks $T$, nor about how or in which order the layers were sampled. In the small example illustrated in Fig.\hs\ref{fig:schematicbenchmark}, generalized modularity \cite{mucha2010community} correctly identifies the communities in each layer but fails to identify the communities in the two original mode networks. Contrarily, the multiplex map equation, here with relax rate $r=0.15$, both identifies the communities in each layer and the communities in the mode networks. 

\begin{figure}[tbhp]
\centering
\includegraphics[trim=40 18 60 40,clip,width=\columnwidth]{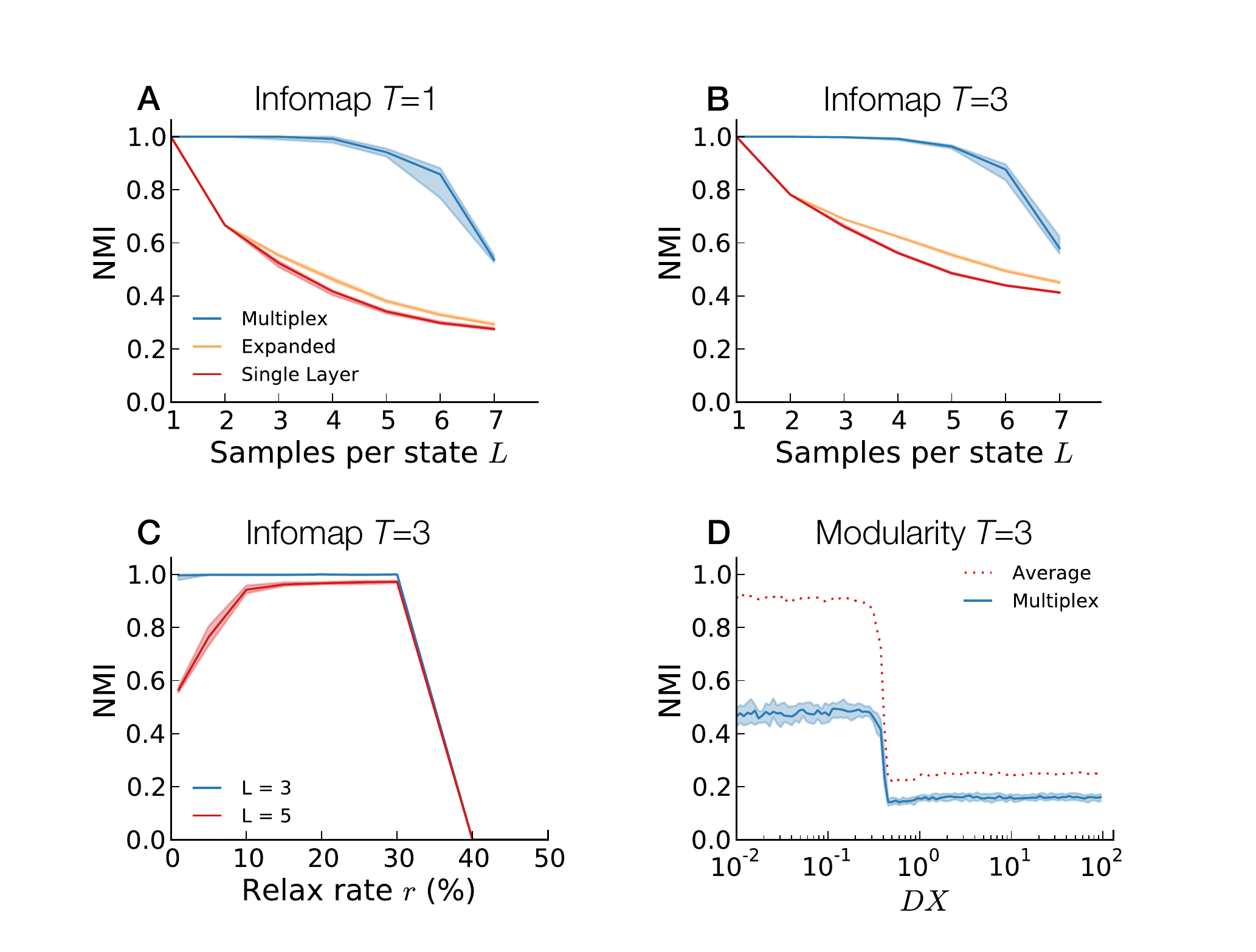}
\caption{\label{fig:benchmarks}Performance test on multilayer benchmark networks. (A-B) Performance of Multiplex Infomap (Multiplex) compared with Infomap applied to the expanded network with state nodes interpreted as physical nodes (Expanded) and to each network layer separately (Single) as a function of number of network layers for 1 and 3 mode networks, respectively. We used relax rate $r=0.15$ and quantified the performance by the NMI between the planted and obtained partitions of state nodes. (C) Performance of Multiplex Infomap as a function of the relax rate $r$. (D) Performance of generalized modularity optimization for $T=L=3$ as a function of the inter-layer coupling $DX$, measured both as the NMI of state nodes (Multiplex) and averaged across network layers (Average). }
\end{figure}

The multiplex map equation can accurately identify multilayer communities.
To test the performance more systematically, we generated multilayer benchmark networks with different number of mode networks $T$ and network layers per mode network $L$. For the mode networks, we used LFR benchmark networks with 128 nodes and 4 communities, each with 32 nodes with average degree 16, and the fraction of inter-community links set to 0.05. We varied $T$ between 1 and 3, and $L$ between 1 and 7. To quantify the performance, we applied the normalized mutual information (NMI) \cite{danon2005comparing, meilua2007comparing} to state nodes. In this way, we can quantify how well the method captures the multilayer communities. Figures\hs\ref{fig:benchmarks}A and B show the results for relax rate $r=0.15$. Optimization of the multiplex map equation with Infomap, Multiplex Infomap for short, accurately identifies the communities of the original mode networks for up to 5-6 network layers per mode network. Contrarily, standard Infomap applied on each layer separately or on the supra-adjacency representation of the multilayer network with all state nodes interpreted as physical nodes \cite{gomez2013diffusion} only succeed for one layer per mode network. That is, only by acknowledging the multiplex nature of the benchmark networks is it possible to accurately identify their multilayer communities. 

The results are only weakly dependent on the relax rate (see Fig.\hs\ref{fig:benchmarks}C), although the exact range depends on the relative constraints on flow manifested in network layers of the same and different mode networks (see Figs.\hs{1} and \hs{2} in the SI). 
When nothing else is stated, we use $r = 0.15$ throughout our analysis. With this relax rate, the random walker stays in the same network layer for about six steps.

Generalized modularity \cite{mucha2010community} does not identify this type of planted communities across network layers (see Fig.\hs\ref{fig:benchmarks}D) because it uses a null model only for intra-layer links and merely a coupling parameter between layers \cite{mucha2010community,petri2014temporal}. As a result, merging different communities across layers always improves the modularity score, as illustrated in Fig.\hs\ref{fig:schematicbenchmark}C.

Overall, we were not able to recover multilayer communities by treating the multilayer network as one large network, as multiple disconnected networks, or as multiple  networks connected with a coupling parameter without a proper null model. 
We conclude that the key discriminating factor is the map equation's ability to capture the important notion of multiplex networks that sets of state nodes across layers represent the very same physical objects.

\begin{figure*}[tbp]
\centering
\includegraphics[width=0.9\textwidth]{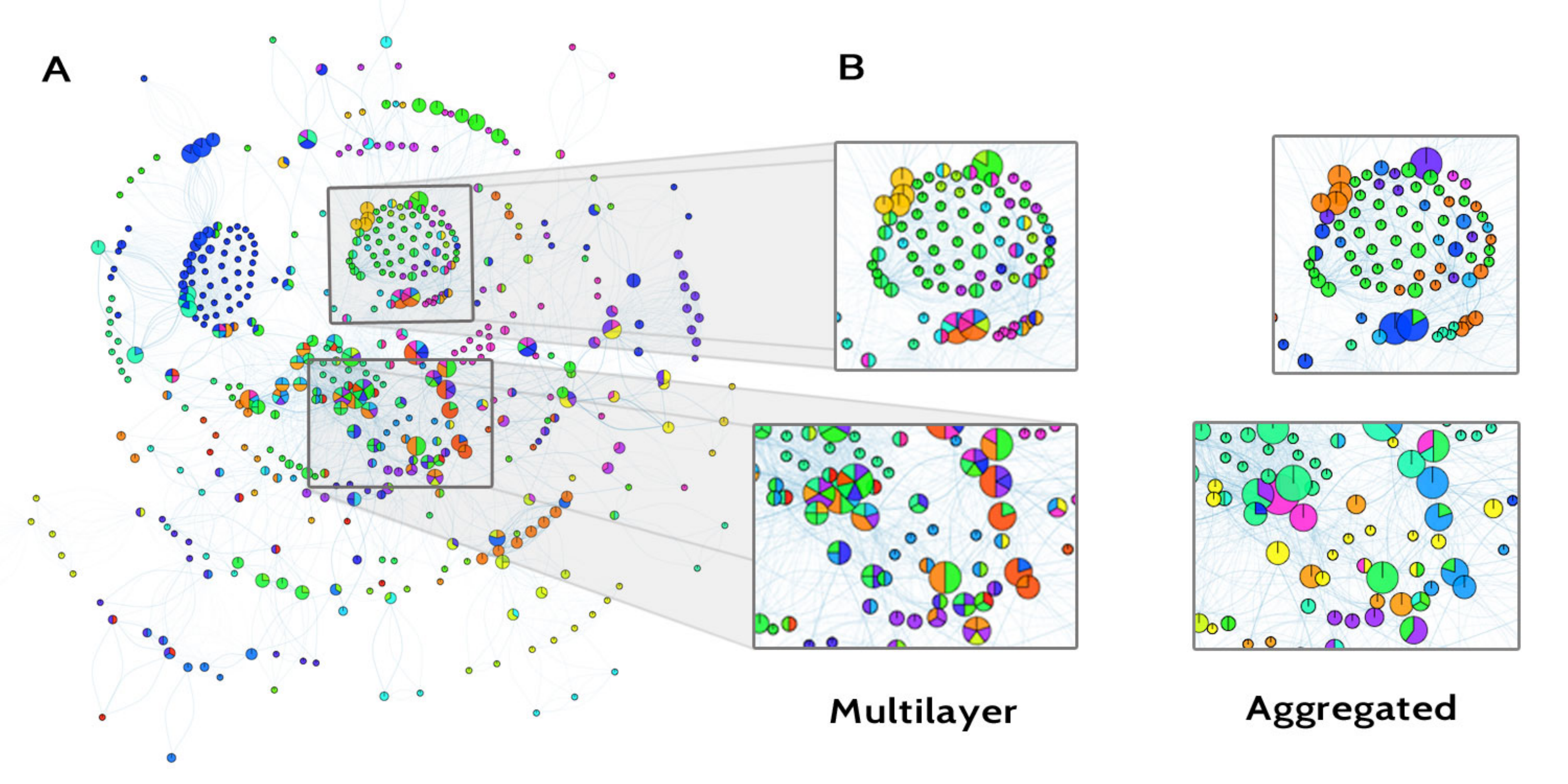}
\caption{\label{fig:auger-partitions-multiplex:main}Community structure in the Pierre Auger Collaboration network. (A) The overlapping community structure revealed by the multiplex map equation with relax rate $r=0.15$. Nodes for scientists, colored according to their module assignments, with node sizes proportional to the number of tasks in which they were active. Specifically, the area of a colored pie-chart slice is proportional to the number of tasks the corresponding scientist is active into. (B) Subsets of nodes with direct comparison with the overlapping community structure obtained from dynamics with $r=1.0$.}
\end{figure*}

\subsection*{Multilayer community structure of collaboration networks} We analyzed two inherently multilayer collaboration networks, the Pierre Auger Collaboration of physicists and a sample from the arXiv of researchers working on networks. The Pierre Auger Collaboration is a group of hundreds of theoretical and experimental scientists worldwide working at the largest observatory of ultra-high energy cosmic rays \cite{pao2013}. The collaborators work together in different research topics on specific tasks, e.g., source detection, mass composition, experimental enhancements, shower reconstruction, etc. Scientists within the Collaboration may work on one or more tasks, and every year hundreds of internal technical reports are submitted to the repository.
With access to author lists and keywords, we reconstructed the inherently multilayer collaboration network in which nodes represent scientists, links indicate collaboration between scientists, and layers represent tasks (see Table\hs{1} in the SI). We considered all submissions between 2010 and 2012, and assigned each report to $L$ layers according to its keywords and its content, with manual disambiguation to avoid spurious results from an automated process. For each report with more than one author, for each layer in which the report was classified, and between any pair of the $N$ co-authors, we added a weight $1/(L(N-1))$ to the weighted, undirected, and multilayer network (see Table\hs{1} for details). In this way, the sum of all link weights of an author across all layers simply is the number of reports written by the author. We built the arXiv multilayer network in the very same way, but instead of tasks we used arXiv categories for layers (see Table\hs{2} in the SI). To restrict the analysis to a well-defined topic of research, we only included papers with ``networks'' in the title or abstract (see Table\hs{1} for details). Because some categories or tasks are more related than others, communities naturally emerge across layers when groups of scientists work on interdisciplinary projects or several tasks simultaneously.

\begin{table}[tbp]
\parbox{1.0\columnwidth}{\footnotesize\raggedright\label{table} Table {{1}}. Summary of multilayer effects on community detection}
\setlength{\tabcolsep}{0pt}
\setstretch{0.85}
\footnotesize{\begin{tabular*}{1.0\columnwidth}{@{}lr@{\hspace{2em}}rrr@{\hspace{2em}}rr}\mytoprule\noalign{\smallskip}
{} && \multicolumn{2}{c}{Synthetic networks} && \multicolumn{2}{c}{Real networks} \\
{} && \multicolumn{1}{c}{$T=1$} & \multicolumn{1}{c}{$T=3$} && \multicolumn{1}{c}{Auger} & \multicolumn{1}{c}{arXiv} \\ 
\cline{0-0}\cline{3-4}\cline{6-7}\noalign{\smallskip}
Number of nodes $n$ && 256 & 256 && 514 & 14,488 \\
Number of links $l$ && 1,400 & 4,000 && 12,964 & 70,350 \\
Number of layers $L_{\mathrm{tot}}$ && 3 & 9 && 16 & 13 \\
\noalign{\smallskip}
NMI, $r_{15}$ vs. $r_{100}$ && 1.0 & 0.0 && 0.74 & 0.92 \\ 
\noalign{\smallskip}
Eff.\ module size, $r_{15}$ && 32 & 11 && 10 & 13 \\
Eff.\ module size, $r_{100}$ && 32 & 128 && 16 & 17 \\
\noalign{\smallskip}
Module assignm., $r_{15}$ && 1.0 & 3.0 && 1.4 & 1.2 \\
Module assignm., $r_{100}$ && 1.0 & 1.0 && 1.1 & 1.0 \\
\noalign{\smallskip}
Persistence gain (\%) && 0 & 163 && 25 & 13\\
Compression gain (\%) && 0 & 32 && 26 & 22 \\
\mybottomrule
\end{tabular*}
}
\parbox{1.0\columnwidth}{\footnotesize\raggedright Synthetic networks with $L=3$ layers per state $T$.
Modeled dynamics denoted $r_{15}$ and $r_{100}$ for relax rate 0.15 and 1.0, respectively.
Effective module size measured as $n/2^{H(\mathcal{S})}$, where $H(\mathcal{S})$ is the entropy of the distribution of module sizes in terms of their flow volumes.
Persistence and compression gains for dynamics modeled with $r_{15}$ compared with $r_{100}$, and with modular solution obtained for $r_{15}$.
All figures are significant.}
\vspace{-0.2cm}
\end{table}

\begin{figure}[tbp]
\centering
\includegraphics[width=\columnwidth]{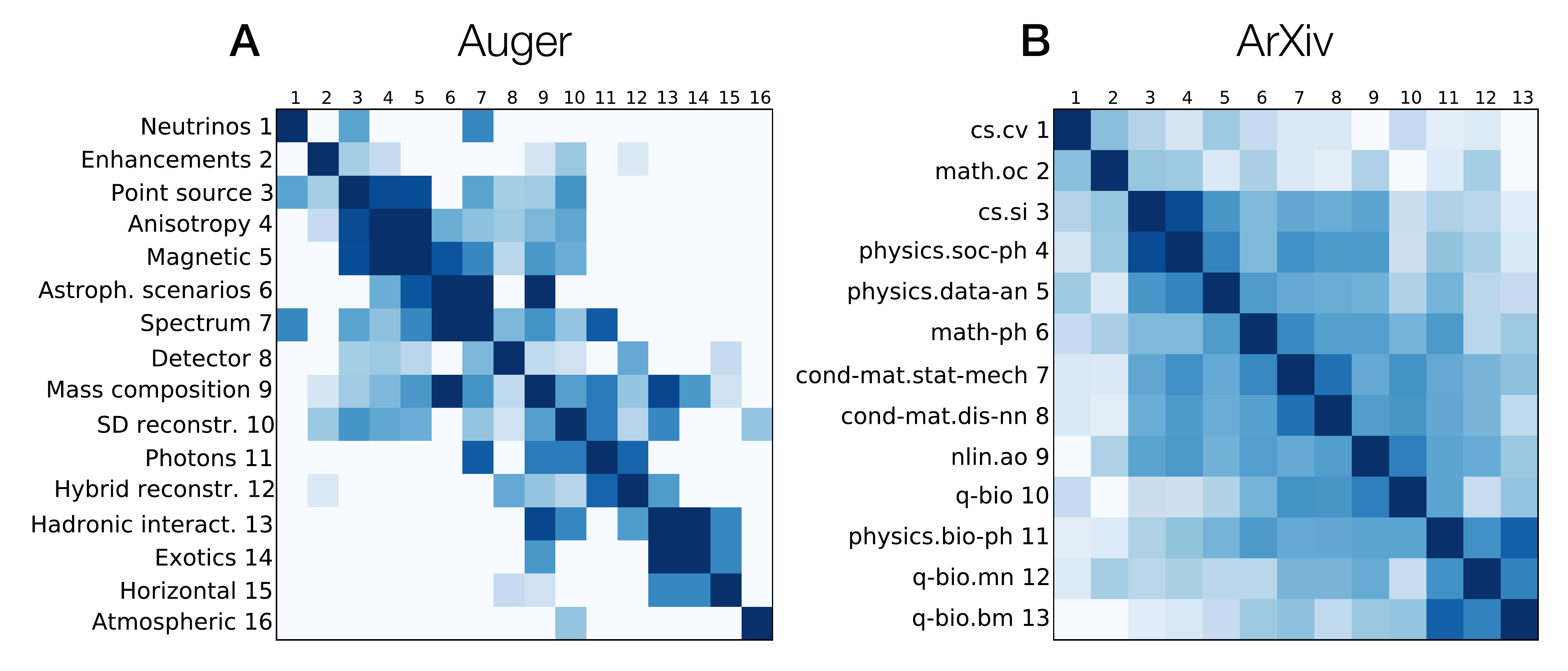}
\caption{\label{fig:layersimilarity} Real multilayer networks with communities across network layers. The heat maps show the similarities between network layers, measured as the fraction of state nodes in different network layers that are assigned to the same communities.}
\end{figure}

The collaboration networks show a highly overlapping modular organization. In Fig.\hs\ref{fig:auger-partitions-multiplex:main}A, we show the largest connected component of the Auger network, including more than 90\% of the scientists, and their assignments into highly overlapping modules. Truly multilayer nodes, i.e., those ones corresponding to scientists active in more than one task, dominate the core of the network in this visualization, whereas single-task scientists are more peripherals nodes. For example, the multilayer analysis reveals strong groups of collaboration across the tasks of ``point source,'' ``anisotropy,'' and ``magnetic,'' (see Fig.\hs\ref{fig:layersimilarity}A). Fig.\hs\ref{fig:auger-partitions-multiplex:main}B shows that essential information about the overlapping modular organization is washed out when dynamics are modeled with $r=1.0$ or in the aggregated network (not shown in the figure because, qualitatively, it provides the same results), and scientists are assigned to a few overlapping communities ($r=1.0$) or one community only (aggregated network).
Without mentioning their names here, we find scientists who indisputable are active in several different tasks with variegated collaboration patterns captured only when dynamics are modeled with $r=0.15$, whereas for $r=1.0$ the scientists are grouped in single non-overlapping communities. In another case, we find two colleagues who work at nearby institutions within the same city and with highly overlapping interests and collaborations. For $r=0.15$, they are assigned to highly overlapping modules across tasks, whereas for $r=1.0$, they are assigned to different non-overlapping partitions.
Only by maintaining the multilayer structure were we able to reveal the actual collaboration structure. Similarly, Fig.\hs\ref{fig:layersimilarity}B shows that communities extend across layers also in the arXiv collaboration network. Whereas communities typically only extend a few layers in the Auger network, communities in the arXiv network can extend over multiple layers. This means that scientists are rather task specific in the Pierre Auger Collaboration, whereas researchers working on networks often are involved in interdisciplinary projects, although computer vision and mathematics seem to be less interdisciplinary topics. In any case, the multilayer networks analyzed with the map equation captures that scientists can simultaneously work in different groups on different topics.  

Table\hs{1} summarizes the multilayer effects of community detection with the map equation framework. For easy comparison, we contrast multilayer results obtained with dynamics modeled with relax rate $r=0.15$, with results obtained with relax rate $r=1.0$ (see Fig.\hs{3} in the SI for full comparison). The latter maximum relax rate corresponds to completely washed out multilayer information, but, unlike the aggregated networks, it allows Multiplex Infomap to assign nodes to multiple modules. For both the Auger and the arXiv networks, we find that flow are confined in smaller and more overlapping modules. We also measure this effect in terms of the persistence gain in modules. For modules obtained with $r=0.15$, the persistence gain quantifies how much longer a random walker stays within the modules with dynamics modeled with $r=0.15$ compared with $r=1.0$. When a random walker only moves freely between layers in one step out of about six compared with free movements between layers in any step, we find that its chance to stay within the same module increases by 25 and 13 percent in the Auger and arXiv network, respectively. As a result of this persistence gain, the modular description of a random walker's trajectory can be significantly compressed in both networks. Since compressing data is dual to finding regularities in the data \cite{shannon1948mathematical,rissanen1978modeling}, the multiplex map equation applied to the multilayer representation allows us to discover patterns that are absent in the aggregated network. Evidently, these patterns contain essential information about the constraints on flow through the systems.

In summary, compared with conventional network analysis, Multiplex Infomap applied to the studied multilayer networks reveals smaller modules with more overlap that better capture the actual organization. Shoehorning multiplex networks into conventional community-detection algorithms can obscure important structural information and earlier attempts of generalizing conventional community-detection methods to identify communities across layers have proven difficult. In contrast, thanks to the map equation's intrinsic ability to capture that sets of nodes across layers represent the very same physical objects in multiplex networks, the framework generalizes straightforwardly. In absence of empirical inter-layer links, here we have modeled the dynamics between layers. However, inter-layer interaction data would provide further important information about the organization of social and biological systems, and calls for more empirical work.

\section*{Methods}
\subsection*{Dynamics on multilayer networks} The rationale behind the multiplex map equation is simple: encode the trajectory between physical nodes of a random walker that itself navigates between state nodes in different layers (see Fig.\hs\ref{fig:schematicmultiplex}C). For a modular description, the trajectory is encoded with unique codewords on all modules and all physical nodes within each module, respectively. We are only interested in the codelength and can derive them from the stationary distribution of the random walker. The stationary distribution on the state nodes can be derived from the transition probabilities $\mathcal{P}_{ij}^{\alpha\beta}$ described in Eq.\hs(\ref{eq:interconnectedtransitionprob}) for interconnected networks with empirical inter-layer link weights and in Eq.\hs(\ref{eq:multilayertransitionprob}) for multilayer networks with inter-layer link weights modeled with relaxation parameter $r$. Assuming that the stationary distribution of state node $i,\alpha$ is $p_{i}^{\alpha}$, it can in principle be derived from the recursive system of equations
\begin{align}
p_{i}^{\alpha} = \sum_{j,\beta} p_{j}^{\beta} \mathcal{P}_{ji}^{\beta\alpha}.\label{eq:stationaryundir}
\end{align}
However, to guarantee a unique ergodic solution in directed networks, we use \emph{teleportation} at a low rate $\tau$ to state nodes proportional to their intra-layer in-strength. \cite{lambiotte2012ranking}. To reduce the smoothening effect of teleportation and make the results more robust to the teleportation parameter $\tau$, we use \emph{unrecorded} teleportation steps and \emph{recorded} steps along links \cite{lambiotte2012ranking}. We obtain the recorded visit rates by first calculating the stationary distribution with teleportation to state nodes proportional to their out-strength,
\begin{align}
\widetilde{p}_{i}^{\alpha} = (1-\tau)\sum_{j,\beta} p_{j}^{\beta} \mathcal{P}_{ji}^{\beta\alpha} + \tau \frac{s_{i}^{\alpha}}{\sum_{i,\alpha}s_{i}^{\alpha}},\label{eq:stationaryrecdir}
\end{align}
with the power-iteration method \cite{golub2012matrix}. Then we derive the recorded steps along links $q_{ji}^{\beta\alpha}$ and nodes $p_{i}^{\alpha}$ in a subsequent step
\begin{align}
q_{ji}^{\beta\alpha} &= \widetilde{p}_{j}^{\beta} \mathcal{P}_{ji}^{\beta\alpha} \\
p_{i}^{\alpha} &= \sum_{j,\beta} q_{ji}^{\beta\alpha}.
\end{align}
We use teleportation rate $\tau=0.15$ throughout our analysis of directed networks, but the results are robust to variation of $\tau$ in a wide range around this value. For undirected networks, results are independent of $\tau$.

\subsection*{The multiplex map equation} We seek to minimize the description length $L({\mathsf M})$ given by the the map equation over possible network partitions $\mathsf M$, with each state node $i,\alpha$ assigned to a module $\boldsymbol{\imath}=1,2,\ldots,m$. The network partition that gives the shortest description length best captures the community structure with respect to the dynamics on the multilayer network. The map equation uses $m$ \emph{module codebooks} and one \emph{index codebook} to describe the random walker's movements within and between modules, respectively, and weights them by how often they are used. From Shannon's source coding theorem \cite{shannon1948mathematical}, the average description length of each codebook is given by the entropy $H(\cdot)$ of the associated probability distribution of events. Therefore,   
both the description length and the rate of use of each codebook can be expressed in terms of the visit rates of the state nodes $p_{i}^{\alpha}$ and the transition rates at which the random walker enters $q_{\boldsymbol{\imath}\curvearrowleft}$ and exits $q_{\boldsymbol{\imath}\curvearrowright}$ each module.
\begin{align}
q_{\boldsymbol{\imath}\curvearrowleft} &= \sum_{i,\alpha \in \boldsymbol{\jmath} \ne \boldsymbol{\imath}, j,\beta \in \boldsymbol{\imath}} q_{ij}^{\alpha\beta}\\
q_{\boldsymbol{\imath}\curvearrowright} &= \sum_{i,\alpha \in \boldsymbol{\imath}, j,\beta \in \boldsymbol{\jmath} \ne \boldsymbol{\imath}} q_{ij}^{\alpha\beta}.
\end{align}
Module codebook $\boldsymbol{\imath}$ has one codeword for all state nodes of each physical node assigned to the module and one exit codeword. The codeword lengths are derived from the rates at which the random walker visits each of the physical nodes in the module, 
\begin{align}
p_{i \in \boldsymbol{\imath}} = \sum_{i,\alpha \in \boldsymbol{\imath}} p_{i}^{\alpha}, \label{eq:physnodeinmod}
\end{align}
and exits the module, $q_{\boldsymbol{\imath}\curvearrowright}$. We use $p_{\boldsymbol{\imath}\circlearrowright}$ to denote the sum of these rates, and $\mathcal{P}^{\boldsymbol{\imath}} = \{ p_{i \in \boldsymbol{\imath}} / p_{\boldsymbol{\imath}\circlearrowright}\}$ to denote the normalized probability distribution. Similarly, the index codebook has codewords for module entries. The codeword lengths are derived from rates at which the random walker enters each module, $q_{\boldsymbol{\imath}\curvearrowleft}$. We use $q_{\curvearrowleft}$ to denote the sum of these rates, and $\mathcal{Q} = \{ q_{\boldsymbol{\imath}\curvearrowleft}/q_{\curvearrowleft} \}$ to denote the normalized probability distribution. We want to express average length of codewords from the index codebook and the module codebooks weighted by their rates of use. Therefore, the map equation is
\begin{align}
L(\mathsf{M}) = q_{\curvearrowleft} H(\mathcal{Q}) + \sum_{\boldsymbol{\imath}=1}^{m}p_{\boldsymbol{\imath}\circlearrowright}H(\mathcal{P}_{\boldsymbol{\imath}}). \label{mapeq}
\end{align}
This is the standard formulation of the map equation \cite{rosvall2008maps} with one important difference: state nodes of a physical node can be assigned to multiple modules, but if they are assigned to the same module they are assigned a common codeword derived from their total visit rate given by Eq.\hs(\ref{eq:physnodeinmod}).

\begin{acknowledgments}
A.A. and M.D.D. were supported by the European Commission FET-Proactive project PLEXMATH (Grant No. 317614) and the Generalitat de Catalunya 2009-SGR-838. A.A. also acknowledges financial support from the ICREA Academia and James S.\ McDonnell Foundation. M.R.\ was supported by the Swedish Research Council grant 2012-3729. The authors acknowledge all members of the Pierre Auger Collaboration for kindly providing access to the meta-data of its repository for internal technical reports, Dr. M. Settimo for kindly helping to classify all reports to the proper task(s), P.J.\,Mucha, M.A.\,Porter, M.\,Bazzi and L.\,Jeub for fruitful discussions.
\end{acknowledgments}

\renewcommand{\figurename}{Figure}
\renewcommand{\thefigure}{S\arabic{figure}}
\renewcommand{\tablename}{Table}
\renewcommand{\thetable}{S\arabic{table}}
\renewcommand{\theequation}{S\arabic{equation}}

\renewcommand{\thesection}{S\arabic{section}}

\setcounter{figure}{0}
\setcounter{table}{0}
\setcounter{section}{0}

\bigskip \textbf{\LARGE Supplementary Information}\\ \bigskip

\section*{OUTLINE}

The supporting information is organized into two sections. First, we provide additional information about the
synthetic benchmark graphs. In particular, we study the impact of the relax rate $r$ on a special set of multilayer networks with different fraction of overlapping communities across layers. Moreover, we provide additional details about the definition of Normalized Mutual Information for multilayer networks. Finally, in Second, we provide additional information about the real datasets that we study. We have made code available at \href{http://www.mapequation.org/}{www.mapequation.org}

\section*{Community structure of synthetic multilayer networks}
\label{sec:benchmarks}

In this section we provide additional information about the synthetic benchmark graphs and associated performance tests.

\subsection*{NMI for multilayer networks}

As explained in the main text, we use Normalized Mutual Information to assess the similarity of the multilayer partitions.
Let us call $X$ the community assignments of one partition and $Y$ the ones of the other partition.
The community assignments are the clusters that the node-layer tuples belong to. 

If we draw a tuple at random (with uniform probability), the probability of observing a certain community $x$ is proportional 
to the number of tuples assigned to it: $p(X=x) = n_x / N $, where $n_x$ is the number of tuples in community $x$ and $N$ is the total
number of tuples. We can also define the joint probability $p(x,y)$, which is proportional to the number of tuples assigned to 
community $x$ in one partition  and community $y$ in the other.

For the NMI, we used the following definition:

\begin{equation}
\label{eq:nmi}
\textrm{NMI} = \frac{H(X) + H(Y) - H(X,Y)} {\max{ \big ( H(X), H(Y) \big )}},
\end{equation}

where $H$ is the Shannon entropy. 

As a final remark, in our synthetic benchmark graphs, when the number of layers is high, it can happen
that some tuples are never sampled. As a consequence, some tuples are in the reference partition but cannot be in
the partition returned by the algorithm. To resolve this issue, we only consider the tuples that appear in both partitions.

\subsection*{Parameters for the LFR benchmark graphs}

Here we give further details about how the LFR benchmark networks \cite{lancichinetti2008benchmark} have been generated.
As mentioned in the main text, our synthetic multiplexes comprises $T$ independent LFR benchmark networks, which we call modes. 
Each of these networks is generated specifying a certain number of parameters, which set the degree distribution, the community sizes and 
how many links are inter and intra the communities. Inside each community, the links are randomly inserted 
via the configuration model \cite{molloy1995critical}, and the 
same model is used to insert the links between the communities.

In the main text, we showed the results obtained for $128$ nodes and $4$ communities of $32$ nodes.
The degree of all nodes was set equal to $16$,
and the mixing parameter, i.e. the average fraction of inter-community links, was set to $5\%$.
These parameters are the same of the popular GN benchmark \cite{girvan2002community}. However, we also 
analysed networks of $1,000$ nodes with heterogenous degree and community size distribution, and found
the same qualitative behaviour.

\subsection*{Benchmarks with overlapping communities across layers} 

We consider a set of synthetic networks where each layer consists of 256 nodes grouped in 8 clique communities poorly interconnected with each other. The communities in all layers are assigned in order to obtain a specific fraction of overlapping nodes across layers, i.e., sub-sets of nodes connected with each other in different layers. In our numerical experiments, we consider different realizations with overlapping fraction ranging from 0 to 1. Moreover, we consider different values of the relax rate, ranging from 0 to 1, to understand how the interplay between structure (i.e., overlapping fraction) and dynamics (i.e., relax rate) affects the detection of the planted partitions.

\begin{figure}[h!]
\begin{center}
\includegraphics[width=3.5in]{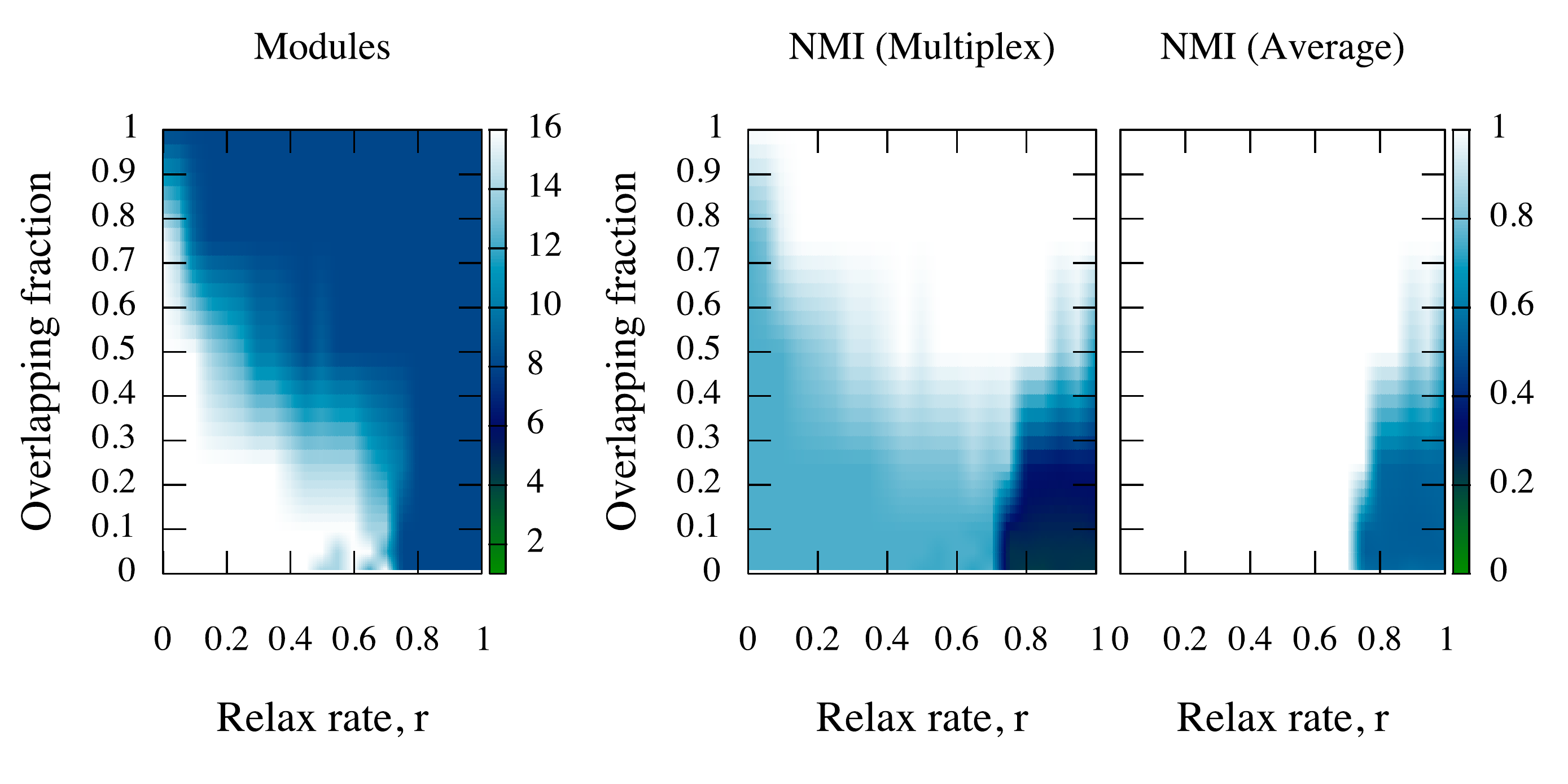}
\caption{\label{fig:bench-overlap}Benchmarks on synthetic multilayer networks. Phase diagrams reporting how the number of modules and the corresponding NMI (Multiplex and Average) change with the relax rate and the fraction of overlapping communities across layers.}
\end{center}
\end{figure}

In Fig.\,\ref{fig:bench-overlap}, we show the resulting phase diagrams for a network with two layers, reporting the number of modules and the resulting NMI of state nodes (Multiplex) and averaged across network layers (Average) against both overlapping fraction and relax rate. First, it is interesting to note that for a wide range of small overlapping fractions and relax rates, the number of detected modules is 16. For overlapping fraction smaller than 50 percent, the networks in the two layers are significantly different. For relax rate lower than 50--60 percent, the layers do not couple and the flow stays preferentially in the cliques within each layer separately. In this region, the two layers behave like if they are not part of the same multiplex and the NMI (Multiplex) estimator is able to detect this behavior, as shown in the right panel of the figure. Conversely, the average NMI suggests that the found partitions per layer are correct: this is equivalent to perform the standard community detection in each layer separately. Outside this region, the two layers are more coupled, because the relax rate is sufficient to allow more information to flow within them and, regardless of overlapping fraction, the number of detected modules is 8. 

For overlapping fraction larger than 50\%, the NMI (Multiplex) is 1 almost regardless of the relax rate. The average NMI is also 1 in the same range, suggesting that the two measures are equivalent in absence of multilayer communities as defined in the main text (see Fig.\,2) and in presence of high overlapping fraction across layers. This result is easily understood in terms of persistent flows across layers and shows that for these multilayer networks, with this specific topology, the average NMI is a more suitable indicator of similarity between the planted partitions and the detected ones.

Moreover, regardless of the overlapping fraction of the underlying multilayer network, 8 modules are always revealed from the analysis of the aggregated network. This result shows the limitations of community detection in aggregated networks: only when the two layers have highly overlapping partitions is the aggregated network a good proxy for the whole multilayer structure. However, this is not the case for the majority of empirical multilayer networks and aggregation could cause a significant loss of information, often leading to a misleading partitioning of the network.

\subsection*{Effect of the relax rate}

For multilayer benchmark networks considered in the main text, the optimal range of relax rates depends on the number of mode networks. For example, with only one mode network and relax rate $r=1$, it is possible to accurately identify the communities in the mode network for any number of network layers (see Fig.~S3). 
For more than one mode network, on the other hand, too high relax rate washes out the constraints set by each mode network, whereas too low relax rate overstates the constraints set by each network layer. Without access to actual inter-layer link weights, the relax rate should be chosen appropriately for the system under study. 

Here we show the NMI for the synthetic benchmarks as a function of the relax rate for a single-mode system: $T=1$. We already showed the same diagram for a three-mode system, $T=3$, in Fig.~3 in the main paper (also shown here for comparison).
If there is a single mode, the optimal solution is achieved for high values of the relax rate, because the data are aggregated.
However, as already shown, aggregating the data cannot find the correct partition if multiple modes are present.

\begin{figure}[h!]
\begin{center}
\includegraphics[width=\columnwidth]{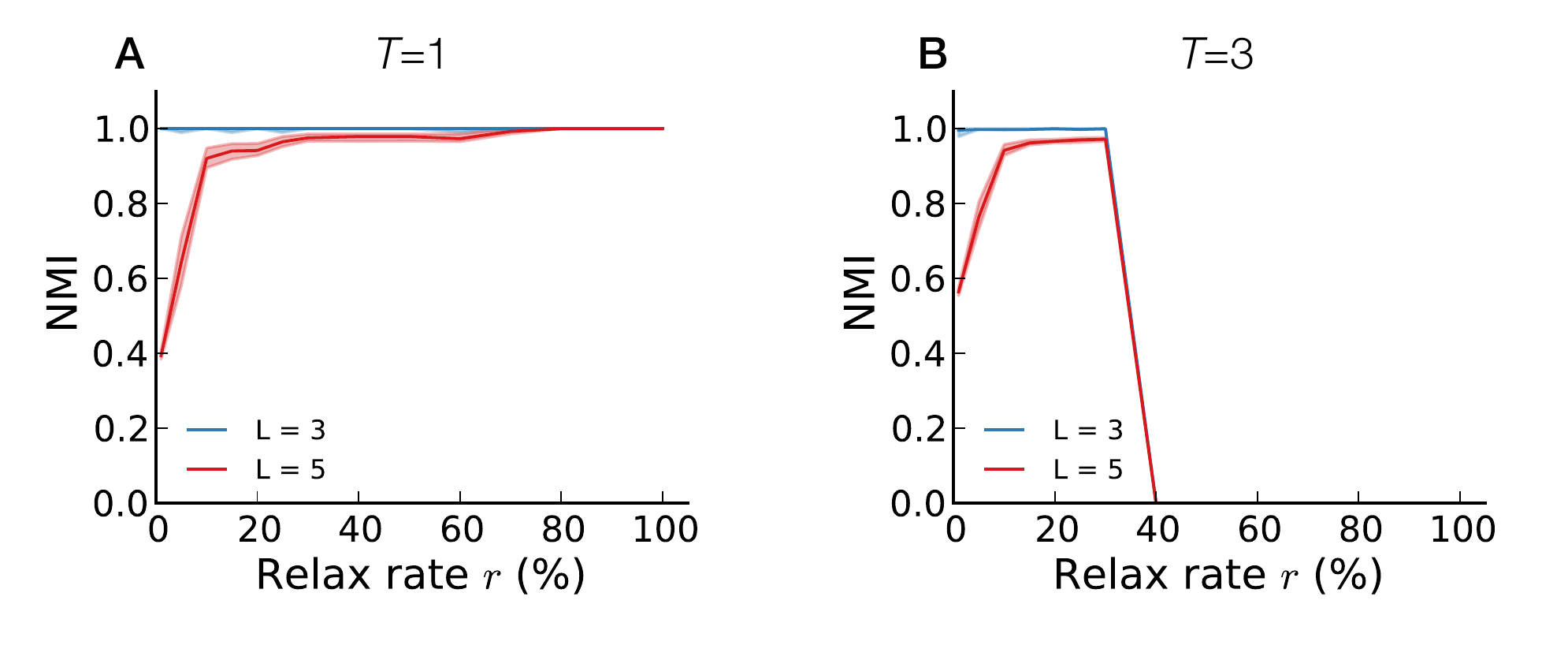}
\caption{NMI for the synthetic benchmark graphs as a function of the relax rate $r$. (A) Performance on a single state system: higher relax rates have a better
performance because are similar to the aggregated. (B) For a three state system, only the multilayer solution can detect the correct partition, as the aggregated returns a single module.}
\label{fig:relaxrate}
\end{center}
\end{figure}

\subsection*{Modularity Optimization}

We performed the same performance tests on generalized modularity \cite{mucha2010community} and Fig.~3D in the main text shows that optimizing modularity can only accurately identify communities in each network layer separately in this specific benchmark test. For no value of inter-layer coupling $DX$ did the method accurately identify the multilayer communities of the mode networks. Specifically, for $DX < 1$, the average NMI shows that the method is capable of detecting the correct partition of each layer, but the multiplex NMI shows that it is not capable of identifying which network layers correspond to which mode networks. For $DX > 1$, the performance drops further, because inter-layer link weights dominate over intra-layer link weights. We conclude that generalized modularity does not identify this type of multilayer communities, because it uses a null model only for intra-layer links and merely a coupling parameter between layers \cite{mucha2010community,petri2014temporal}. As a result, merging different communities across layers always improves the modularity score, as illustrated in Fig.~3C.

\section*{Comparing the aggregated and multilayer networks of Auger and ArXiv }
\label{sec:realdata}

\begin{figure}[h!]
\begin{center}
\includegraphics[width=\columnwidth]{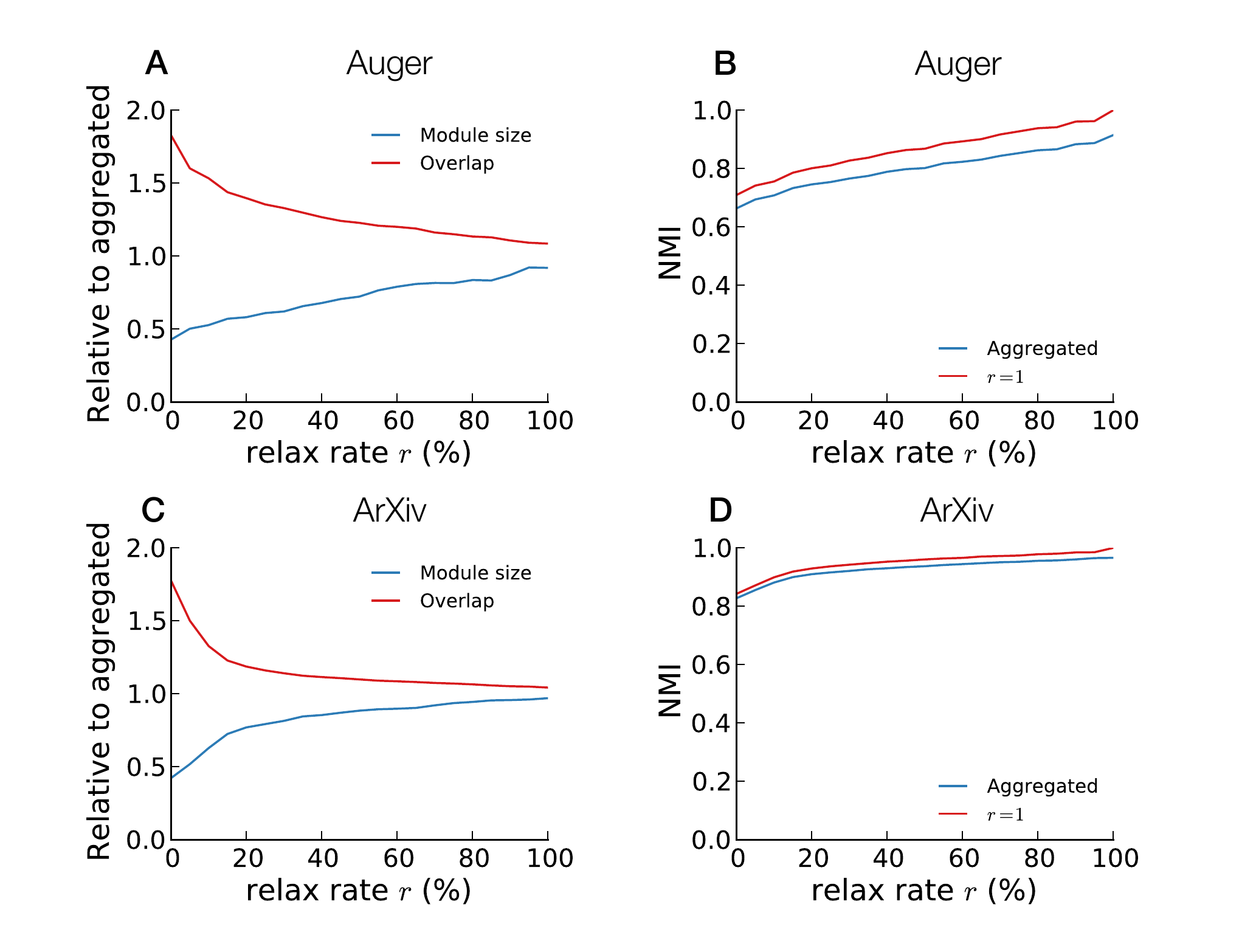}
\caption{\label{fig:auger-plots} \textbf{Aggregation is responsible for significant changes in the community structure.} Module sizes, number of assignments per node (overlap) and NMI for communities revealed from the multilayer and the aggregated networks in the Pierre Auger Collaboration (top panels) and the ArXiv (bottom panels) networks.
For a given relax rate, the NMI measures the similarity between the obtained partition and the partitions obtained from the aggregated topology (blue curve) and the aggregated dynamics at relax rate $r=1.0$ (red curve), respectively.
}
\label{fig:compare-real}
\end{center}
\end{figure}

Figure~\ref{fig:compare-real} shows how differences between the partitions found with the multilayer and the aggregated approach. At increasing relax rate, the random walker becomes less and less localized in a specific layer. Accordingly, the NMI between the multilayer and the aggregated solutions increases. For $r=1$, the walker moves freely between layers, but the NMI is not one because the multilayer
solution still allows for overlap (the optimization algorithm we used on the aggregated does not identify overlaps by construction).

In both datasets, with increasing relax rate, we find bigger modules (module size increases) and fewer community assignments per physical node (overlap decreases).
The module size is defined as the number of nodes divided by the effective number of modules. The effective number of modules is defined as $2^H$, where $H$ is the Shannon entropy of the partition (see previous Section).

\subsection*{The Pierre Auger Collaboration Dataset}

We considered all internal technical reports submitted to the Pierre Auger \footnote{Official web page of the Pierre Auger Observatory \protect\url{http://www.auger.org}. Note that the GAP repository is not publicly accessible.} GAP repository  between 2010 and 2012, assigning each report to one or more tasks according to its keywords and its content and performing disambiguation manually, to avoid unavoidable spurious results due to automated processes. The final number of distinct authors in this dataset is 514, with 9,209 collaborations, classified in 16 layers. Finally, we built the international co-authorship network for each layer, excluding only those papers with only one author to avoid self-loops in the corresponding adjacency matrices. Privacy policies have been considered and we anonymized the data by assigning a random numerical integer to each author. The list of layers is shown in Table\,\ref{tab:auger} together with the corresponding tasks.

\begin{table}[!t]
\caption{\label{tab:auger} Pierre Auger Observatory: each task defines a layer in the multilayer co-authorship network}
\begin{small}
\begin{tabular}{clcl}
\\\hline
\textbf{Layer ID} & \textbf{Task} & \textbf{Layer ID} & \textbf{Task}\\\hline
1 & Neutrinos & 9 & Spectrum\\
2 & Detector & 10 & Photons \\
3 & Enhancements & 11 & Atmospheric \\
4 & Anisotropy & 12 & SD Reconstr. \\
5 & Point Source & 13 & Hadronic Interact.\\
6 & Mass Composition & 14 & Exotics \\
7 & Horizontal & 15 & Magnetic \\
8 & Hybrid Reconstr. & 16 & Astroph. Scenarios \\
\end{tabular}
\end{small}
\end{table}

\subsubsection*{Detailed community structure.} 
\begin{figure*}[!b]
 \centering
\includegraphics[width=18cm]{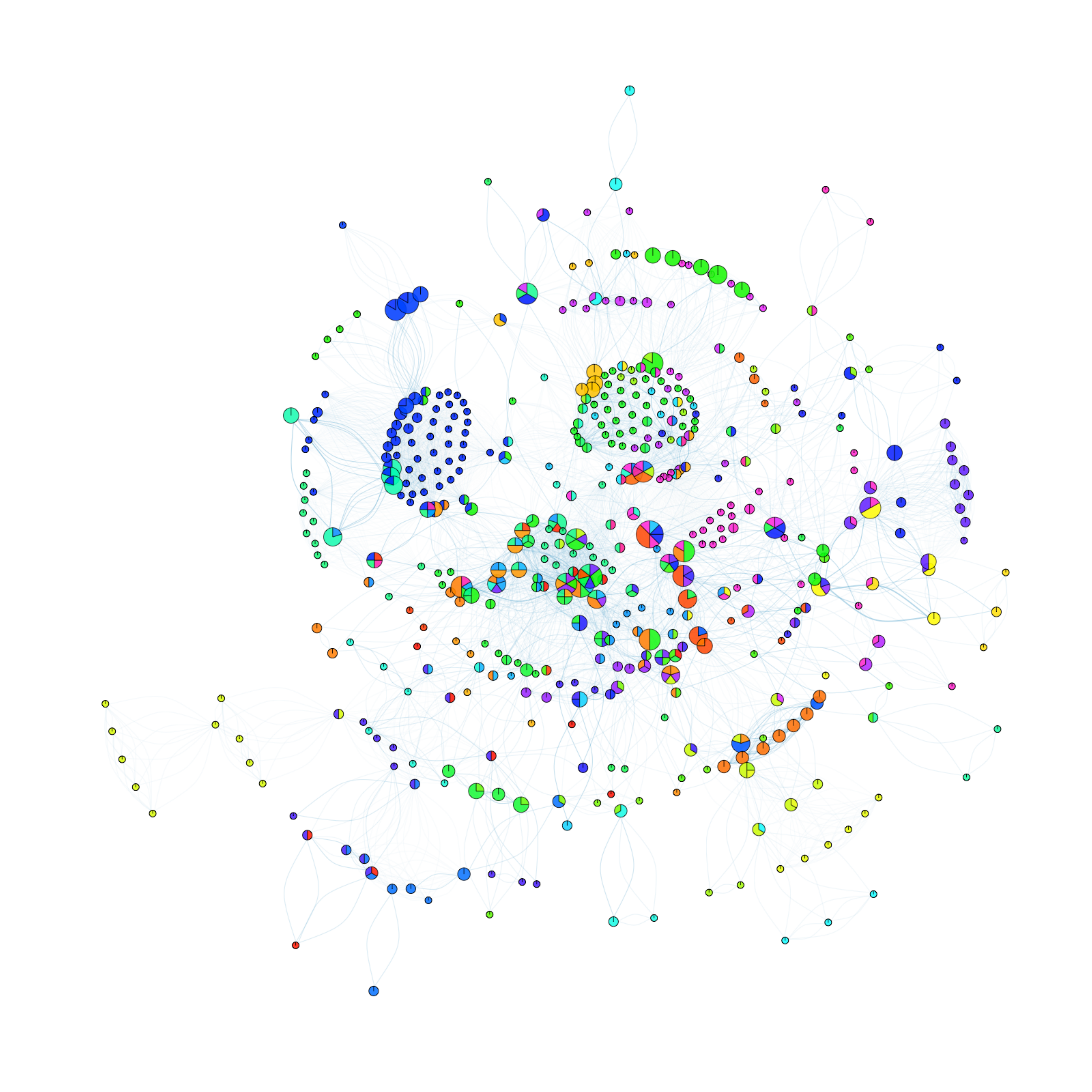}
 \caption{\label{fig:auger-partitions-multiplex}\textbf{Community detection in the Pierre Auger Collaboration network.} Detailed map of the partitions obtained applying the map equation with relax rate $r=0.15$ to the multilayer network. The size of a node is proportional to the multilayer activity of the corresponding author: larger the number of tasks where he or she collaborates larger the size of the node. Colors within the pie chart code the different communities where the author is classified into, where the area of each slice is proportional to the number of times the author is classified in the corresponding community.}
 \end{figure*}

\begin{figure*}[!b]
 \centering
\includegraphics[width=18cm]{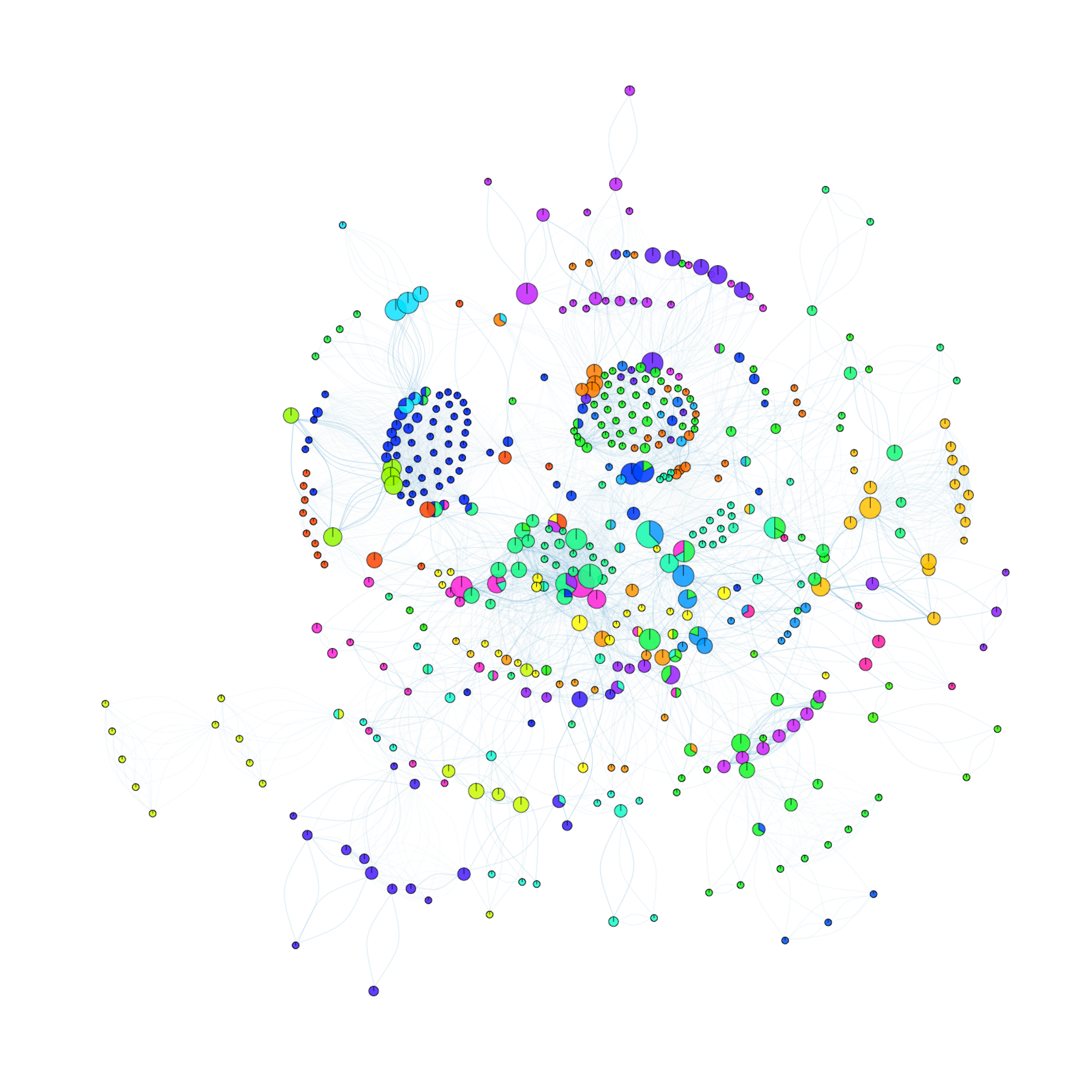}
 \caption{\label{fig:auger-partitions-aggregate}\textbf{Community detection in the Pierre Auger Collaboration network.} As in Fig.\,\ref{fig:auger-partitions-multiplex}, considering $r=1.0$.}
 \end{figure*}

We show in Fig.\,\ref{fig:auger-partitions-multiplex} and Fig.\,\ref{fig:auger-partitions-aggregate} more detailed maps of the community structures shown in the main text for the multilayer and the aggregated networks, respectively.

\clearpage

\subsection*{The ArXiv Collaboration Dataset}

We queried the ArXiv\footnote{Official web page of the ArXiv repository \protect\url{http://www.arxiv.org} }, the free and publicly accessible repository of scientific pre-prints, at the end of May 2014. We focused our search on the 13 categories listed in Table\,\ref{tab:arxiv}, defining the layers of the multilayer co-authorship network, and we collected all papers containing the word ``network'' either in the title or in the abstract. We found 12,019 articles between 1,993 and May 2014 from 14,488 authors whose names have been heuristically disambiguated. We assigned each pre-print to the corresponding categories and we created a link between two authors if they co-authored a paper.

\begin{table}[!t]
\caption{\label{tab:arxiv} ArXiv repository: each category defines a layer in the multilayer co-authorship network}
\begin{small}
\begin{tabular}{clcl}
\\\hline
\textbf{Layer ID} & \textbf{Category} & \textbf{Layer ID} & \textbf{Category}\\\hline
1 & physics.soc-ph & 8 & q-bio.MN\\
2 & physics.data-an & 9 & q-bio \\
3 & physics.bio-ph & 10 & q-bio.BM \\
4 & math-ph & 11 & nlin.AO \\
5 & math.OC & 12 & cs.SI\\
6 & cond-mat.dis-nn & 13 & cs.CV \\
7 & cond-mat.stat-mech & & \\
\end{tabular}
\end{small}
\end{table}

\bibliographystyle{unsrtnat}
\bibliography{multimap}

\end{document}